\documentclass[a4paper,final,oneside,notitlepage,12pt]{article}
\usepackage[T1]{fontenc}
\usepackage{amsfonts}
\usepackage{amssymb}
\usepackage{amsthm}
\usepackage{amsmath}
\usepackage[all]{xy}
\usepackage{amsfonts}
\usepackage{amssymb}
\usepackage{amsmath}
\usepackage{color}
\usepackage{graphicx}

\newtheorem{theor}[subsection]{Theorem}
\newtheorem{theo}[subsubsection]{Theorem}
\newtheorem{rem}[subsubsection]{Remark}
\newtheorem{defi}[subsection]{Definition}

\newtheorem{lemma}[subsubsection]{Lemma}

\def\al{\alpha}

\def\calc {\mathcal{C}}

\def\ka{\kappa}
\def\la{\lambda}
\def\La{\Lambda}

\def\lf{\lfloor}
\def\nn{\nonumber}
\def\one {{\mathchoice {\rm 1\mskip-4mu l} {\rm 1\mskip-4mu l}
{\rm 1\mskip-4.5mu l} {\rm 1\mskip-5mu l}}} 

\def\rf{\rfloor}
\def\si{\sigma}
\def\Si{\Sigma}

\makeatletter

\long\def\@makecaption#1#2{
 \vskip\abovecaptionskip
 \sbox\@tempboxa{{\bf #1}: {#2}}
 \ifdim \wd\@tempboxa >10cm
   \begin{list}{}{
     \leftmargin=15mm
     \labelwidth=0mm
     \rightmargin=15mm
     \listparindent=0mm
     \itemsep=0mm
   }
   \item{\bf #1}: {#2}\end{list}
 \else
   \global \@minipagefalse
   \hb@xt@\hsize{\hfil\box\@tempboxa\hfil}
 \fi
 \vskip\belowcaptionskip}

\def\proof {{Proof.}\hspace{7pt}}
\def\endofproof {\hfill{$\Box$}\\}

\def\adress#1{\gdef\@adress{#1}}
\def\@adress{}
\def\preprint#1{\gdef\@preprint{#1}}
\def\@preprint{}
\def\maketitle{
 \newpage
 \noindent
 \begin{tabular}{cc}
   \begin{minipage}[c]{0.4\textwidth}
     \begin{flushleft}
      \includegraphics[width=110pt]{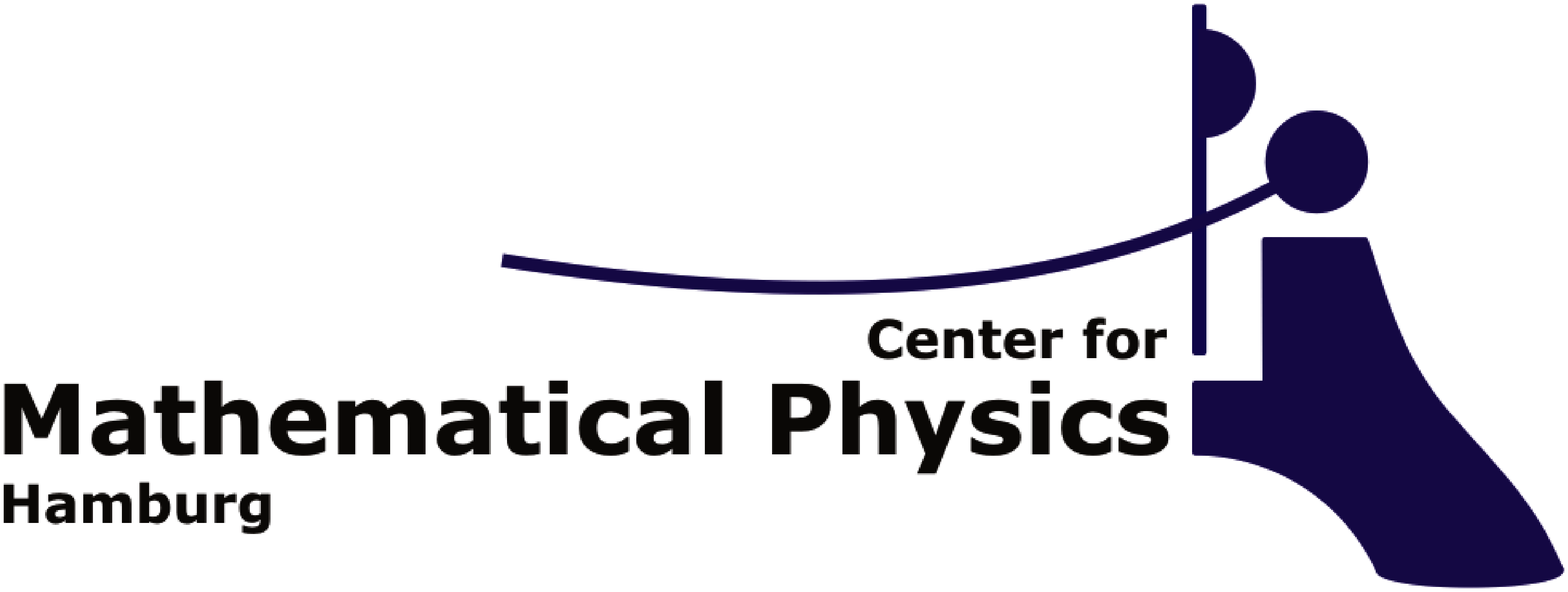}
     \end{flushleft}
   \end{minipage}&
   \begin{minipage}[c]{0.6\textwidth}
     \begin{flushright}
     {\small\sf\@preprint}
     \end{flushright}
   \end{minipage}
 \end{tabular}
 \vskip 3cm
 \begin{center}
   \LARGE\@title
   \if!\@author!\else \vskip 0.5cm \large\@author\fi
   \if!\@adress!\else \vskip 0.5cm \normalsize\@adress\fi
 \end{center}
 \vskip 2cm
}

\makeatother

\begin{document}

\title{Kramers-Wannier dualities for WZW theories and 
minimal models}

\author{Christoph Schweigert and Efrossini Tsouchnika}

\adress{Organisationseinheit Mathematik\\Schwerpunkt Algebra und
Zahlentheorie\\Universit\"at
Hamburg\\Bundesstra\ss e 55\\D--20146 Hamburg}

\preprint{arXiv:hep-th/0710.0783\\
Hamburger Beiträge zur Mathematik Nr.\ 276 \\
ZMP-HH/07-8}

\maketitle

\begin{abstract}
We study Kramers-Wannier dualities for Wess-Zumino-Witten theories
and (super-)minimal models in the Cardy case, i.e.\ the case with
bulk partition function given by charge conjugation. Using the
TFT approach to full rational conformal field theories,
we classify those dualities that preserve all chiral symmetries. 
Dualities turn out to exist for small levels only. 
\end{abstract}

\thispagestyle{empty}

\newpage
\setcounter{page}{1}

\section{Introduction}

Two-dimensional conformal field theories have been an essential tool to study 
universal properties of critical phenomena. 
They capture surprisingly many aspects of statistical models with critical 
points, even away from criticality.
A fascinating aspect of some of these models are
Kramers-Wannier like dualities, relating e.g.\ the high-temperature and low-temperature regime; the critical point is typically self-dual. 
While this has been known for more than sixty years \cite{krwa},
the obvious question whether such dualities can be deduced from properties 
at the critical point has only been addressed recently.

The -- affirmative -- answer uses 
\footnote{See also \cite{ruel} for a different approach.}
an algebraic approach to full (rational) 
conformal field theories \cite{fuRs,fuRs4} that describes 
correlation functions of these theories in terms of two pieces of data: \\
$\bullet~~$ the chiral data of the conformal field theory, which are
encoded in a  modular tensor category $\calc$ \\
$\bullet~~$ a (symmetric special) Frobenius algebra $A$ in the tensor
category $\calc$. 

For the purposes of the present paper, a modular tensor category $\calc$ 
(see \cite{tur} and e.g.\ \cite{baki}) is defined to be a semi-simple 
$\mathbb C$-linear 
abelian ribbon category with simple tensor unit $\one$, having a 
finite number of isomorphism classes of simple objects; the braiding 
on the tensor category is required to
obey a certain nondegeneracy condition. (This definition is slightly more 
restrictive than the original one in \cite{tur}.) 

In the TFT approach to rational conformal field theory, types of topological 
defect lines correspond to isomorphism
classes of $A$-bimodules. Given two bimodules $B_1$ and $B_2$, their
tensor product $B_1\otimes_A B_2$ is again a bimodule; this tensor product 
encodes the fusion of topological defects.
In the same way the modular tensor category $\calc$ of chiral data gives 
rise to a fusion ring $K_0(\calc)$ of chiral data, the tensor category 
$\calc_{AA}$ of $A$-bimodules gives rise to a fusion ring $K_0(\calc_{AA})$ 
of topological defects. Both fusion 
rings are semi-simple. The fusion ring of defects is, however, not
necessarily commutative, since the category of bimodules is typically 
not braided. (For a more detailed discussion of the fusion
ring $K_0(\calc_{AA})$ see e.g.\ \cite{fuRs12}.)

The present paper builds on the insight of \cite{ffrs3,ffrs4} that
the fusion ring $K_0(\calc_{AA})$ of topological defects 
determines both symmetries and Kramers-Wannier dualities of the full conformal
field theory described by the pair $(\calc,A)$. The fact that Kramers-Wannier 
dualities relate \cite{kace} bulk fields to disorder fields located at the 
end points of topological defect lines might have suggested a relation between
Kramers-Wannier dualities and topological defects.
However, the topological defect lines relevant for the dualities are 
of a different type than the ones created by the dual disorder fields.

We explain the pertinent results of \cite{ffrs3,ffrs4} in more detail:
symmetries of the full conformal field theory $(\calc,A)$ correspond to 
isomorphism classes of invertible objects in \ $\mathcal{C}_{AA}$. These are 
objects $B$ satisfying

\begin{equation}
B \otimes_A B^\vee \cong A \,, \,\,\,\,
B^\vee \otimes_A B 
\cong A 
\,,
\end{equation}
where  $B^\vee$ is the bimodule dual to $B$.
The isomorphism classes of the invertible bimodules form a group,
the Picard group $Pic(\mathcal{C}_{AA})$. This group is not necessarily 
commutative: for the three state Potts model, for example, it turns out to be
the symmetric group $S_3$ on three letters, see \cite{ffrs3}.

To describe dualities, we need the following

\begin{defi}
Given a modular tensor category $\cal C$ and a simple symmetric special 
Frobenius algebra $A$ in $\calc$, a simple $A$-bimodule $B$ is called 
a {\em duality bimodule} \cite{ffrs3} iff all simple subobjects of the
tensor product $B^\vee \otimes_A B$ are invertible bimodules.
\end{defi}
It is easy to see that for any duality bimodule $B$ the isomorphism classes of 
simple bimodules $B_\lambda$ 
such that $\dim_{{\mathbb C}} \mathrm{Hom}(B_\lambda, B^\vee \otimes_A B) > 0$ 
form a subgroup $H$ of the Picard group $Pic( {\cal C}_{AA})$ and that
in this case $\dim_{\mathbb{C}} \mathrm{Hom}(B_\lambda, B^\vee \otimes_A B)=1$.
We call $H$ the {\em stabilizer} of the duality bimodule $B$.

In the present paper, we restrict ourselves to the case where
the simple symmetric special Frobenius algebra in $\cal C$ is the
tensor unit $\one$; this situation is usually referred to as the
Cardy case. A complete set of correlation functions for the full conformal
field theory in the Cardy case has been constructed in \cite{fffs2}. 
In the Cardy case, the bulk partition function is given by charge conjugation. 
Many more simplifications occur: 
The category of $A$-bimodules is equivalent to the 
original category, $\mathcal{C}_{AA}\cong \cal C$; as a consequence,
isomorphism classes of invertible bimodules are simple currents \cite{scya6}:

\begin{defi}
(i)~~A {\em simple current} in a modular tensor category $\cal C$ is an 
isomorphism class $[J]$ of simple objects $J$ satisfying 
\begin{equation}
J \otimes J^\vee \cong \one \,. 
\end{equation}
(ii)~~ A {\em fixed point} $[U_\phi]$ of a simple current $[J]$  is an 
isomorphism class of simple objects of $\cal C$ satisfying
\begin{equation}
J \otimes U_\phi \cong U_\phi \,. 
\end{equation}
(iii) A {\em duality class} is an isomorphism class $[U_\phi]$ of simple 
objects such that the tensor product $U_\phi^\vee\otimes U_\phi$ is isomorphic 
to a direct sum of invertible objects containing at least two non-isomorphic
invertible objects.
\end{defi}
As a special case of the results of \cite{ffrs3,ffrs4}, we see 
that the full conformal field theory in the Cardy case
has Kramers-Wannier dualities if and only if the underlying 
modular tensor category has duality classes.

In the present paper, we study two classes of unitary rational
conformal field theories: (super-)Virasoro minimal models and 
Wess-Zumino-Witten (WZW) theories.
There is a WZW theory for every reductive 
finite-dimensional complex Lie algebra; in the present paper, we limit 
ourselves to simple Lie algebras. Thus, both classes of conformal field
theories come in families that are parametrized by a positive integer,  
called the level. 

Simple currents in Virasoro minimal models (and thus their symmetries) are 
well known; for WZW theories, simple currents have been classified in \cite{jf15}: with the 
exception of the WZW theory based on $E_8$ at level 2, they are in bijection 
with the center of the corresponding simple, connected, simply-connected 
compact Lie group. It might be surprising at first sight that only the
center - rather than at least the full Lie group -- shows up as the
symmetry group. One should keep in mind, however, that the symmetries
we discuss are required to preserve all chiral symmetries, i.e.\
the complete current algebra. Relaxing this requirement -- which amounts
to working with the representation category of a subalgebra  of the
current algebra and a nontrivial symmetric special Frobenius algebra
in this category -- leads to larger symmetry groups.

In this paper, we present a classification of Kramers-Wannier dualities
in these models. To state our results, we need to introduce some notation:
In WZW theories the chiral algebra is generated by an untwisted affine Lie
algebra $X^{(1)}_r$ of rank $r$; moreover, one has to fix a positive integral
value $k$ of the level.  We denote the corresponding modular tensor
category by ${\mathcal C}[ (X_r)_k]$.

In this paper we establish the following results:

\begin{theor} 
The only Wess-Zumino-Witten theories $({\mathcal C}[(X_r)_k],\one)$ with 
duality classes are $ (E_7)_2,(A_1)_2, (B_r)_1$ and $(D_{2r})_2$ with
$r\geq 2$. 
\end{theor}

Since the finite-dimensional complex simple Lie algebras $B_2$ and $C_2$
are isomorphic, we do not list $(C_2)_1$ separately.
It should be noted that dualities only appear at low level. We are,
unfortunately, not aware of any a priori argument for this finding. 
For all cases except $(D_{2r})_2$, the stabilizer is cyclic of order two.
For most of the cases, the existence of the dualities does not come as a 
surprise: the modular tensor categories for $(A_1)_2$ and $(B_r)_1$ 
have the same fusion rules as the Ising model. For $(E_7)_2$ the
fusion rules are isomorphic to those of the tricritical Ising model and
hence the category contains a tensor subcategory with Ising fusion rules.

Concerning Virasoro minimal models and their superconformal
counterparts, we find the following situation.

\begin{theor}
(i)~~ Duality classes only exist for the unitary Virasoro minimal 
models at level $k=1$ and $k=2$, i.e.\ for the Ising model and
the tricritical Ising model. \\
(ii)~~ Duality classes for an $N=1$ super-Virasoro 
unitary minimal model only exist for odd levels. If they exist, they are
unique. \\
(iii)~~ $N=2$ superconformal minimal models have duality classes only for level $k=2$.
\end{theor}

Our findings agree, for the nonsupersymmetric theories, with the ones of
\cite{ruel}; our methods, however, are different. Again we find a close 
relation to
Ising fusion rules: all models with a Kramers-Wannier duality have a 
realization by a coset construction involving $(A_1)_2$.

\medskip

The plan of the paper is as follows: Section 2 contain model independent
results. We analyze properties of duality classes; in Theorem \ref{ddc} we 
give a necessary condition for the existence of duality classes that is
used in the analysis of the $A$-series of Wess-Zumino-Witten models and the 
minimal models.
 
For the $B$-, $C$-, $D$-series and the exceptional algebras
$E_6$ and $E_7$, there are no simple objects meeting the conditions 
of Theorem \ref{ddc}. A different strategy involving lower bounds of quantum 
dimensions  is developed in Section 2; it is based on the results
of \cite{garw} on second-lowest quantum dimensions in Wess-Zumino-Witten
fusion rules. Section 3 contains the analysis of dualities for
Wess-Zumino-Witten theories; Section 4 is devoted to the study of 
(super-)conformal minimal models.

\section{Model-independent considerations}

\subsection{Preliminary remarks and notation}

We start by introducing some notation: we choose a set 
$(U_\lambda)_{\lambda\in I}$ of representatives for the isomorphism classes 
of simple objects of 
the modular tensor category $\calc$. In particular, given $\mu\in I$,
we find a unique $\overline\mu$ such that $U_\mu^\vee \cong U_{\overline\mu}$.
The tensor unit $\one$ of a modular tensor category is simple; we choose the 
representatives such that $U_0=\one$ and $0\in I$.

The classes $[U_\lambda]$ form the distinguished basis of the fusion ring 
$K_0(\calc)$. The  fusion coefficients
$$ {\cal N}_{\la \mu}^{\ \rho}:=\dim_{\mathbb{C}} \mathrm{Hom}(U_\la \otimes 
U_\mu , U_\rho) $$
are the structure constants of the multiplication on $K_0(\calc)$.
The following identities are
easy consequences of the properties of a duality of a tensor category:
\begin{equation}
\label{fu}
{\cal N}_{\la \mu}^{\ \rho}={\cal N}_{\mu\, \overline\rho}^{\ \overline\la}
={\cal N}_{\overline\rho\la}^{\ \ \overline\mu}\,. 
\end{equation}

The braiding isomorphism $c_{U,V}: U\otimes V \to V\otimes U$
of a modular tensor category furnishes a symmetric matrix 
$s_{ij} = \mathrm{tr} c_{U_j,U_i}\circ c_{U_i,U_j}$. 
For the models considered in this paper, there exists a positive real
factor such that the matrix $s$ rescaled by this factor is unitary.
We call this unitary matrix $S$. The fusion coefficients can be expressed
in terms of the modular matrix $S$ by the Verlinde formula
\begin{equation}
\label{v}
{\cal N}_{\rho \si}^{\ \tau} = \sum _{\kappa\in I} 
\frac{S_{\rho \kappa}S_{\si \kappa}\bar 
S_{\tau \kappa}}{S_{0 \kappa}}\ .
\end{equation}
Another property of the modular matrices $S$ in the models of our interest
are the inequalities
$S_{0\ka}\geq S_{0 0}> 0$ for any simple object $U_\ka$. Entries
for dual objects are related by complex conjugation,
$S_{\lambda\overline\mu}= \overline{S_{\lambda\mu}}$.
We finally note the following easy consequence of the Verlinde formula 
(\ref{v}) and the unitarity of $S$  
\begin{equation}
\label{v2}
\frac{S_{\rho \kappa}}{S_{0 \kappa}}
\frac{S_{\si \kappa}}{S_{0 \kappa}}
 = \sum _{\tau\in I} {\cal N}_{\rho \si}^{\ \tau}\ 
\frac{S_{\tau \kappa}}{S_{0 \kappa}}\ \,\,\ .
\end{equation}

\medskip

We need to introduce two more notions for modular tensor categories.
Recall the definition of a simple current from the introduction. 
The set of simple currents carries the structure of a finite abelian group, 
called the Picard group $Pic( {\cal C})$.
It turns out that every isomorphism class $[U_\ka]$ of simple objects of
$\calc$ gives rise to a character on the group $Pic( {\cal C})$:
\begin{equation} \label{mon}
\chi_\la([U_\ka]):= \frac{S_{\la \ka}}{S_{0 \la}} \,\, ,
\end{equation}
which we call the monodromy character of the object $U_\ka$.

\medskip

A modular tensor category being in particular a ribbon tensor
category, there is the notion of a (quantum) dimension for any object
$U$ of $\calc$. It depends only on the isomorphism class of an
object. For simple objects, we introduce the abbreviation
$$ {\cal D}_{\la}:= \dim (U_\lambda)\, . $$
The quantum dimension is related to the modular $S$-matrix via
\begin{equation}
{\cal D}_{\la}=\frac{S_{0 \la}}{S_{0 0}}\ .
\end{equation}

The quantum dimension is a ring homomorphism from $K_0(\calc)$ to the ring
of algebraic integers over the rational numbers. Dual objects
have identical dimension
\begin{equation}
\label{pr}
{\cal D}_{\overline\la}={\cal D}_{\la}\, .\nn
\end{equation}
Since the quantum dimensions of the modular tensor categories we consider are
the Frobenius-Perron eigenvalues of their fusion matrices, they
are real and obey ${\cal D}_{\la}\geq 1$. Equality is achieved
precisely for simple currents.

It follows immediately from the properties of the quantum dimension 
that the quantum dimension of a duality object of $\calc$ 
is the square root of an integer
${\cal D}_{\phi}=\sqrt{|H|}$, with $|H|$ the order of the stabilizer
$H\leq Pic(\cal C)$.

\medskip

To discuss WZW theories, we finally need some Lie-theoretic notation.
Let $X^{(1)}_r$ be an untwisted affine Lie algebra. Denote by $\La^i$ the 
fundamental weights and by $a_i^\vee$ the dual Coxeter labels. 
Their sum equals the dual Coxeter number, $g^\vee=\sum_{i=0}^r a_i^\vee$.
A labelling of the nodes of the Dynkin diagram provides a labelling
of simple roots and fundamental weights; we use the conventions of
\cite{KAc3,FUch}.

Fix a nonnegative integer $k$, the level. At level $k$, there are finitely 
many integrable highest weights $\la$ 
\begin{equation}
\label{hw}
\la\in P_+^k=\{(\la_0, \la_1, \dots \la_r):=\sum_{i=0}^r \la_i\La^i\  
|\ \la_i\in\mathbb{Z}_{\geq0},\ \sum_{i=0}^r a_i^\vee\la_i=k\}\,.\nn
\end{equation}
Isomorphism classes of simple objects of the modular tensor category
${\mathcal C}[ (X_r)_k]$ are in bijection to elements of $P_+^k$;
in particular, the irreducible highest weight representation with
highest weight $k\La_0$ is the tensor unit $U_0$.

For fixed level $k$, the zeroth component of a highest weight is redundant. 
We therefore work with the finite-dimensional simple Lie algebra $\bar X_r$,
called the horizontal subalgebra, and the horizontal part 
$\bar{\la}=\sum _{i=1}^r\la _i \La ^i$ of the weight $\la$.
The quantum dimension of the simple object $U_\la$ can be expressed by a deformed
version of Weyl's dimension formula in terms of a product over a set of 
positive roots of the horizontal subalgebra $\bar X_r$:

\begin{equation}
\label{qd}
{\cal D}_{\la}^{\ (k)}=\prod_{\bar{\al} > 0} 
\frac{\lf (\bar{\la}+ \bar\rho , \bar{\al})\rf_k}
{\lf (\bar\rho ,\bar{\al})\rf_k}\ .
\end{equation}
Here $\bar\rho=\sum_{i=1}^r\La^i$ is the Weyl vector of $\bar X_r$ and for
given level $k\in \mathbb{N}$, the bracket 
$\lf x \rf_k$ of a rational number $x$ is the real number
\begin{equation}
\lf x\rf_k:=\sin \left( \frac{\pi x}{k+g ^\vee }\right) \, .\nn 
\end{equation}
The identity
\begin{equation}
\lf x \rf_k = \lf k+g^\vee-x \rf_k\ .
\end{equation}
is immediate.

\subsection{A criterion for the existence of dualities}

We start with the following useful criterion:

\begin{theo} \label{ddc}
Let $H\le Pic(\cal C)$ be a subgroup of simple currents.
Suppose that there is a simple object $U_\mu$ with the following
two properties:
\begin{enumerate}
\item The restriction of the monodromy character $\chi_\mu$ of $U_\mu$
to $H$ is nontrivial.
\item The tensor product of $U_\mu$ with its dual object contains, apart from
the tensor unit, just one more simple object:
$$U_\mu \otimes U_\mu^{\ \vee}\cong \one \oplus U_\nu \,\, . $$
\end{enumerate}
Then a duality class with stabilizer $H$ can only exist if the simple
object $U_\nu$ appearing in the tensor product $U_\mu \otimes U_\mu^{\ \vee}$
is a non-trivial simple current.
\end{theo}

\proof 
Suppose $U_\phi$ is a duality class; then
formula (\ref{v2}) gives for any $\kappa\in I$:
\begin{equation*}
\label{v3}
\left\vert \frac{S_{\phi \kappa}}{S_{0\kappa}} \right\vert ^2
=\frac{S_{\phi \kappa}}{S_{0\kappa}}
\frac{S_{\overline\phi \kappa}}{S_{0\kappa}}
=\sum _{J\in H}\frac{S_{J \kappa}}{S_{0\kappa}}\ \,\, , 
\end{equation*}
where $H$ is the stabilizer of $U_\phi$.

The definition of the monodromy character (\ref{mon}) together with
standard properties of characters of finite abelian groups implies
\begin{equation} \label{tr}
\left\vert \frac{S_{\phi \kappa}}{S_{0\kappa}} \right\vert ^2
=\sum_{J\in H} \chi_\kappa(J) 
\, \, . 
\end{equation}
This expression is nonvanishing iff the restriction of the monodromy 
character of $U_\kappa$ to $H$ is trivial; in this case it equals the
order $|H|$ of the group $H$.
Thus $S_{\phi\ka}=0$, whenever $U_\phi$ is a duality class and
$U_\kappa$ a simple object whose monodromy character restricted to $H$
is nontrivial.

The second property of the simple object $U_\mu$ together with
(\ref{v2}) immediately gives
\begin{equation*}
\left\vert \frac{S_{\mu\phi}}{S_{0\phi}} \right\vert^2=
1+\frac{S_{\nu\phi}}{S_{0\phi}}\ .
\end{equation*}
By the first assumption, $U_\mu$ has nontrivial monodromy
character; therefore $S_{\mu\phi}=0$. As a consequence,
\begin{equation}
\label{p1}
1+\frac{S_{\nu\phi}}{S_{0\phi}}=0\, .
\end{equation}

Next, we notice that due to its appearance in the tensor product
$U_\mu\otimes U_\mu^\vee$, the object $U_\nu$ has necessarily
trivial monodromy character. Therefore, relation (\ref{tr}) yields
\begin{equation*}
\left\vert \frac{S_{\phi \nu}}{S_{0\nu}} \right\vert ^2
= \left\vert H \right\vert\ \mbox{for}\ U_\nu\ \ \mbox{and}\ \ 
\left\vert \frac{S_{\phi 0}}{S_{0 0}} \right\vert ^2
=\left\vert H \right\vert\ \mbox{for the tensor unit}\ U_0\,.
\end{equation*}
Taking the quotient of the last two relations gives
\begin{equation}
\left\vert \frac{S_{\phi \nu}}{S_{\phi 0}} \right\vert ^2
=\left\vert \frac{S_{0 \nu}}{S_{0 0}} \right\vert ^2 
=\mathcal{D}_\nu^{\ 2}\,\, .\nn 
\end{equation}
Equation (\ref{p1}) now implies that the left hand side of this equation 
equals one; hence $\nu$ has to be a simple current.
\endofproof
\begin{rem}
Since $[U_\nu]$ is a simple current, the simple object $U_\mu$ is
itself a duality class for the cyclic stabilizer generated by $[U_\nu]$.
The simple current $[U_\nu]$ has order two.
\end{rem}
The criterion of Theorem \ref{ddc} will be applied to the $A$-series and the
minimal models in Section 3.

\subsection{Monotonicity of quantum dimensions}

Let $[J]$ be a simple current and $\phi\in I$. Due to the relation
$$ {\cal N}_{\phi \overline\phi}^{\ \ \ J} = 
{\cal N}_{J \phi }^{\ \ \ \phi} $$
the simple current $J$ appears in the decomposition of the tensor product 
$U_\phi\otimes U_\phi^\vee$ if and only if $[U_\phi]$ is a fixed point 
of $[J]$.
Duality classes are thus, in particular, fixed points under a
subgroup $H$ of the Picard group.   A fixed point under
a subgroup $H$ of the Picard group is a duality class with stabilizer
$H$ if and only if its 
quantum dimension equals $\sqrt{|H|}$.

To get constraints on the existence of duality classes, we study the growth
of the quantum dimension of fixed points of subgroups of the Picard
group as a function of the level. The following lemma (cf.\ also
\cite{jf18}) plays a key role for finding duality classes for WZW theories 
based on untwisted affine Lie algebras in
the $B$-, $C$-, $D$- series and exceptional Lie algebras.

\begin{lemma}
\label{incr}
Let $X_r^{(1)}$ be an untwisted affine Lie algebra. Let $\la(k)$ be a family 
of integral highest weights of $X_r^{(1)}$ at level $k$ of the form
$$ \lambda(k) = \sum_{i=1}^r \la_i\La^i\ + (k-\sum_{i=1}^r a_i^\vee \lambda_i)
\Lambda^0
\,\, , $$
i.e.\ where only the zeroth component of $\lambda(k)$ depends on the level.
Assume that not all $\la_i$ vanish.
Denote by ${\cal D}_{\la} (k)$ the quantum dimension of $\lambda(k)$
at level $k\in \mathbb{Z}_{\geq 0}$. Then the expression of equation
(\ref{qd}) for ${\cal D}_{\la} (k)$ defines 
an analytic function on $\mathbb{R}_{\geq0}$ which 
is strictly monotonically increasing.
\end{lemma}

\proof
 Differentiating equation (\ref{qd}) with respect to $k$ yields
\begin{equation}
\label{fd}
\frac{\partial}{\partial k}{\cal D}_{\la} (k)={\cal D}_
{\la} (k)\sum_{\bar{\al} >0}(f_k((\rho ,\bar{\al}) )
-f_k((\bar{\la} +\rho ,\bar{\al})))\,,
\end{equation}
where the function 
$$f_k(x):=\frac{\pi x}{(k+g^\vee)^2}\cot (\frac{\pi x}{k+g^\vee})$$
is strictly monotonically decreasing as a function of $x$ for all fixed
levels $k$.
Thus all summands are nonnegative; moreover, $(\bar \la,\bar{\al})> 0$ for at least one positive root
$\bar{\al}$. Hence the expression (\ref{fd}) is positive.
\endofproof

\section{Dualities for WZW theories}

The modular tensor categories based on the Lie algebras $E_8$ for level 
greater or 
equal to three, $F_4$ and $G_2$ do not have non-trivial
simple currents \cite{jf15}. 
As a consequence, no duality classes exist. At level two, $E_8$ has
Ising fusion rules and thus has a duality class with cyclic stabilizer
of order two.

In the sequel we will use the slightly redundant notation 
${\cal D}_{\la}^{(k,r)}$ for the quantum dimension of the weight 
$\la$ of the algebra $(X_r)_k$ at level $k$. Simple objects will
be referred to by the horizontal part of their weight; for reasons of simplicity,
we will drop overlines over horizontal weights.

\subsection{The affine Lie algebra $E_7$}

The tensor categories based on $E_7$ have  cyclic Picard group of
order two, generated by the irreducible highest weight representation
with highest weight $k\La^6$. We are thus led to classify fixed points
of quantum dimension $\sqrt{2}$. 
The action of a simple current on highest weights corresponds to a
symmetry of the Dynkin diagram; the nodes invariant under the symmetry
corresponding to the nontrivial simple current 
all have even Coxeter labels. Since fixed points correspond to highest weights
invariant under this symmetry, they -- and hence duality classes -- only occur 
at even level.

According to \cite{garw}, the second smallest quantum dimension for given level
$k\geq 5$ occurs for horizontal weight $\Lambda^6$,
${\cal D}_{\la} ^{\ (k)}\geq {\cal D}_{\Lambda^6}^{\ (k)}$.
The monotonicity lemma \ref{incr} yields 
\begin{eqnarray}
{\cal D}_{\Lambda^6}^{\ (k)}
\geq {\cal D}_{\Lambda^6}^{\ (5)}=\frac{\lf 10 \rf_5}
{\lf 1\rf_5 } >\ \sqrt{2}. \nn
\end{eqnarray}
Thus, there are no duality classes for level $k\geq 5$.

The simple objects of second lowest quantum dimension at level $k=2,4$
have been listed in Table 3 of \cite{garw}. At level $k=2$, one
finds $\Lambda^7$ which is a fixed point of quantum dimension
$$ {\cal D}_{\La^7}^{\ (2)}=\frac{\lf 2 \rf_2 \lf 6\rf_2
\lf 8\rf_2}{\lf 3\rf_2 \lf 4\rf_2 \lf 5
\rf_2 \lf 7\rf_2}= \sqrt{2} $$
and thus a duality class. Indeed, the fusion rules of $E_7$ at level 2 
are isomorphic to the fusion rules of the tricritical Ising model which
is known to exhibit a Kramers-Wannier duality. 

At level 4, according to the same table, the second smallest quantum dimension 
is assumed for $2\Lambda^7$. The quantum dimension is larger than $\sqrt2$
(this can be shown e.g.\ by using the computer program {\sc kac} \cite{kac}).
We conclude that
WZW theories based on the untwisted affine Lie algebra $E_7^{(1)}$
exhibit a duality class only at level two.

\subsection{The affine Lie algebra $E_6$}
The Picard groups of the tensor categories based on $E_6$ are cyclic of order 
three. Duality objects are 
therefore precisely the fixed points of quantum dimension $\sqrt{3}$. 
It follows from the values of the Coxeter labels that they can only
occur at levels divisible by three. 

According to \cite{garw}, for $k\geq 3$, the second smallest quantum dimension 
occurs for the
weight $\Lambda^1$, ${\cal D}_\la^{\ (k)}\geq {\cal D}_{\Lambda^1}^{\ (k)}$.
By monotonicity in $k$, we derive a lower bound for the quantum dimensions
of simple objects, provided the level $k$ is not smaller than three,
\begin{eqnarray}
{\cal D}_\la^{\ (k)}\geq {\cal D}_{\Lambda^1}^{\ (3)}= 
\frac{\lf3\rf_3\lf6\rf_3}{\lf 1\rf_3\lf 4\rf_3}>\sqrt{3} \nn
\end{eqnarray}
and deduce that there are no Kramers-Wannier dualities for WZW theories
based on $E_6$.

\subsection{The series $C_r$}

The Picard group of WZW theories based on the untwisted affine Lie
algebra $C^{(1)}_r$ is cyclic of order two;
a duality class must be a fixed point of quantum dimension $\sqrt2$.
It follows from the values of the Coxeter labels that the level $k$ must be 
even for odd rank $r$; for even rank there is no restriction on the level.

The isomorphism of complex simple Lie algebras $C_1\cong A_1$ allows us
to assume that $r\geq 2$. For $r\geq 2$ and $k=1$ or $k+r\geq 6$ the 
second minimal quantum dimension is given \cite{garw} by   
\begin{eqnarray}
\label{c2}
{\cal D}_{\la} ^{(k,r)}\ge
{\cal D}_{\Lambda^1} ^{(k,r)}= \frac{\lf \frac{r}{2}
\rf_k \lf r +1\rf_k} {\lf \frac{1}{2}\rf_k \lf 
\frac{r+1}{2}\rf_k}\ .\nn 
\end{eqnarray}
Lemma \ref{incr} states ${\cal D}_{\Lambda^1} ^{(k,r)}
\geq {\cal D}_{\Lambda^1} ^{(1,r)}$ with equality of $k=1$
and we compute
\begin{eqnarray}
{\cal D}_{\Lambda^1} ^{(1,r)}
=\frac{\lf 1\rf_1\lf \frac{r}{2}\rf_1 }
{\lf \frac{1}{2}\rf_1 \lf \frac{r+1}{2}\rf_1} \geq \sqrt2\,,\nn 
\end{eqnarray}
with equality for $r=2$. Indeed, the tensor category for $(C_2)_1
\cong (B_2)_1$ has Ising fusion rules and therefore displays a
Kramers-Wannier duality.

For $k+r<\ 6$ second minimal quantum dimensions are
\begin{eqnarray}
{\cal D}_{2\Lambda^1}^{\ (2,2)} &=& \frac{\lf \frac{5}{2} 
\rf_2 \lf 1\rf_2} {\lf 1\rf_2 \lf 2\rf_2} =2 > \sqrt2  \nn \\
{\cal D}_{\Lambda^1}^{\ (3,2)} &=& \frac{\lf 1 \rf_3 \lf 3\rf_3}
{\lf \frac{1}{2}\rf_3 \lf \frac3 2\rf_3}= 1+\sqrt3 >\sqrt{2}  \nn \\
{\cal D}_{\Lambda^1}^{\ (2,3)} &=& \frac{\lf\frac3 2 \rf_2
\lf 4\rf_2} {\lf \frac{1}{2}\rf_2\lf 2 \rf_2} = 1+\sqrt3
>\sqrt{2}\,.\nn  
\end{eqnarray}
As a consequence, the only duality class occurs for $C_2$ at level 1.

\subsection{The series $B_r$}

Again the Picard group is cyclic of order two so that we are led to
classify fixed points of quantum dimension $\sqrt2$.

Because of the isomorphisms $C_1\cong A_1$ and $C_2\cong B_2$ of 
finite-dimensional complex Lie algebras, we restrict ourselves to $r\geq 3$.
For $r\geq 3$ and $k\geq 4$ or $k=2$ the second minimal quantum dimension
 is given \cite{garw} by   
\begin{eqnarray}
\label{b2}
{\cal D}_{\Lambda^1}^{(k,r)} = \frac{\lf r+\frac{1}{2}\rf_k
\lf 2r-1\rf_k} {\lf 1\rf_k \lf r-\frac{1}{2}\rf_k}\nn
\end{eqnarray}
and with the monotonicity lemma \ref{incr} we obtain
\begin{eqnarray}
{\cal D}_{\Lambda^1}^{(k,r)}
> {\cal D}_{\Lambda^1}^{(2,r)}= \frac{\lf 2\rf_2\lf r+
\frac{1}{2}\rf_2} {\lf 1\rf_2 \lf r-\frac{1}{2}\rf_2}=2\, .\nn
\end{eqnarray}
This excludes duality classes for all levels $k\geq 4$ and $k=2$.

At level $k=1$, we find Ising fusion rules and thus the unique duality
class $\Lambda^r$. At level $k=3$, the second lowest quantum dimension
is assumed for the weight $3\Lambda^r$ \cite{garw}. For its quantum
dimension, we find
$$ {\cal D}_{3\La ^r}^{(3,r)}= \frac{\lf 2\rf_3}{\lf r\rf_3}
\prod_{l=1}^r \frac{\lf 2l-1\rf_3} {\lf l-\frac{1}{2}\rf_3}
= \frac{\lf 2\rf_3}{\lf r\rf_3}
\prod_{l=1}^r \cos(\pi \frac{l-\frac{1}{2}}{2r+2})\ , $$
which  is     strictly larger than $\sqrt2$ for all ranks
$r\geq 2$:
We consider the case of even and odd rank separately; for even
rank we find
$$ {\cal D}_{3\La ^r}^{(3,r)}= \frac{\lf 1\rf_3\lf 2\rf_3}{\lf \frac{1}{2}\rf_3
\lf \frac{3}{2}\rf_3}
\prod_{l=2}^{\frac{r}{2}} \frac{\lf 2l-1\rf_3\lf 2l-1\rf_3} {\lf 2l-\frac{1}{2}\rf_3
\lf 2l-\frac{3}{2}\rf_3}\ \frac{\lf r+1\rf_3}{\lf r\rf_3} \,\, ; $$
with the notation 
$\lceil x\rceil_k:=\cos \left( \frac{\pi x}{k+g ^\vee }\right)$  this
is equal to
$$ {\cal D}_{3\La ^r}^{(3,r)}= \frac{\lf 1\rf_3}{\lf \frac{1}{2}\rf_3}
\frac{\lf 2\rf_3}{\lf \frac{3}{2}\rf_3}
\prod_{l=2}^{\frac{r}{2}} \frac{1-\lceil 4l-2\rceil_3}{\lceil 1\rceil_3-
\lceil 4l-2\rceil_3}\ \frac{\lf r+1\rf_3}{\lf r\rf_3}\ .$$
Since the arguments of the sine- and cosine-functions are all smaller than 
$\pi /2$,
all quotients are bigger than one. The first quotient is strictly larger 
than $\sqrt{3}$, because $\frac{\frac{\pi}{2}}{2r+2}<\frac{\pi}{6}$ for any 
value of $r$. Therefore, the quantum dimension ${\cal D}_{3\La ^r}^{(3,r)}$ is 
strictly larger than $\sqrt2$.

We proceed in an analogous way for odd rank to find
$$ {\cal D}_{3\La ^r}^{(3,r)}= \frac{\lf 1\rf_3}{\lf \frac{1}{2}\rf_3}
\frac{\lf 2\rf_3}{\lf \frac{3}{2}\rf_3}
\prod_{l=2}^{\frac{r-1}{2}} \frac{1-\lceil 4l-2\rceil_3}{\lceil 1\rceil_3-
\lceil 4l-2\rceil_3}\ \frac{\lf r\rf_3}{\lf r-\frac{1}{2}\rf_3}
\geq\sqrt 3\ .$$
We conclude that there are no duality classes for levels $k\geq 2$.

\subsection{The series $D_r$}

The structure of the Picard group depends on whether the rank is odd or 
even. For even rank, it is isomorphic to the Kleinian four group 
$\mathbb{Z}_2\times \mathbb{Z}_2$ with generators
$k\Lambda^{r-1}$ and $k\Lambda^{r}$. Accordingly, we have to
consider this group as well as its three cyclic subgroups of order two
as possible stabilizers. For odd rank, the Picard group is
cyclic of order four with $k\Lambda^{r-1}$ or $k\Lambda^{r}$
as possible generators. The possible stabilizers are then
the full Picard group and the cyclic group of order two generated by
$k\Lambda^1$.

For $r\geq 4$ and $k\geq 3$ the second minimal quantum dimension is given by   
\begin{eqnarray}
\label{d2}
{\cal D}_{\Lambda^1}^{(k,r)}= \frac{\lf r\rf_k\lf 2r-2\rf_k}
 {\lf 1\rf_k \lf r-1\rf_k}\nn
\end{eqnarray}
and a lower bound is by Lemma \ref{incr}
\begin{eqnarray}
{\cal D}_{\Lambda^1}^{(k,r)}
 > {\cal D}_{\Lambda^1}^{(2,r)}=
 \frac{\lf 2\rf_2\lf r\rf_2} {\lf 1\rf_2\lf r-1\rf_2}=2\,.\nn 
\end{eqnarray}

We deduce that for $k\geq 3$ no duality classes for any
stabilizer exist: for cyclic stabilizers of order two this
is immediate. Duality objects whose
stabilizer is the full Picard group must be fixed points under
all four simple currents. They can only occur at even levels; but for
level equal to four and higher, by the monotonicity properties of
the quantum dimensions, no simple object of quantum dimension two
exists.

The remaining case is level $k=2$. It is convenient to
treat $D_4$ separately: in this case, simple objects 
have quantum dimension equal to one or two. There is a single
duality class with highest weight $\Lambda^2$ which is a fixed point under
the whole Picard group. We find a single duality class.

For rank greater or equal to five and level $k=2$, there are four simple 
currents, four simple objects of quantum dimension strictly bigger than two
and all simple objects with weights $\Lambda^1$, $\Lambda^{r-1}+\Lambda^r$
and $\Lambda^i$ with $i=2,\ldots r-2$  have quantum dimension 2.
Among these weights, however, only $\Lambda^{r/2}$ for even rank
is a fixed point under the whole Picard group and provides a duality
class with the Picard group $\mathbb{Z}_2\times \mathbb{Z}_2$ as the
stabilizer.

\subsection{The series $A_r$}

In this case, we can use Theorem \ref{ddc} with $U_\mu$ equal to the defining 
$(r+1)$-dimensional irreducible
representation with highest weight $\Lambda^1$. Monodromy classes of
WZW theories are in correspondence with conjugacy classes of representations.
For $A_r$ the defining representation is a generator of the group of 
conjugacy classes of representations and hence has nontrivial monodromy 
character. For the tensor product with the dual simple object we find
\begin{equation*}
\Lambda^1 \otimes (\Lambda^1)^\vee= \Lambda^1\otimes \Lambda^{r-1}
\cong 0 \oplus (\Lambda^1+\Lambda^r).
\end{equation*} 
The second simple object in the direct sum is an invertible object 
only for $r=1$ and $k=2$. Thus, the only duality class appears for $(A_1)_2$
which is known to have Ising fusion rules.

\section{Dualities for (super-)minimal models}

\subsection{Virasoro minimal models}

Nonsupersymmetric Virasoro minimal models can be obtained by the
coset construction \cite{goko2} from the diagonal embedding
\begin{equation}
(A_1)_{k+1} \hookrightarrow (A_1)_k\oplus (A_1)_1
\ . \nn
\end{equation}
To describe isomorphism classes of simple objects, consider triples 
$\Phi ^{l\ s}_{\ t}$,
where $l\in \{0,\dots k\},\ s\in \{0,1\}$ and $t\in \{0,
\dots k+1\}$, subject to the condition
\begin{equation}
\label{mm0}
l+s-t=0\bmod2 \, . 
\end{equation}
On such triples, simple objects are equivalence classes of the relation
\begin{equation*}
\Phi ^{l\ s}_{\ t}\,\,\sim \Phi ^{k-l\ 1-s}_{\ k+1-t}\, .
\end{equation*}

The decomposition of the tensor product of simple objects into a direct
sum of simple objects is given by 
\begin{equation*}
\Phi ^{l_1\ s_1}_{\ t_1}\otimes \Phi ^{l_2\ s_2}_{\ t_2} \,\, \cong \,\,
\bigoplus_{l_3=|l_1-l_2|}^{l_{\max}}\ \ \bigoplus_{t_3=|t_1-t_2|}
^{t_{\max}}\Phi ^{l_3\ s_3}_{\ t_3}\, 
\end{equation*}
with $l_{\max}=\min(l_1+l_2, 2k-l_1-l_2)$ and 
$t_{\max}=\min(t_1+t_2, 2k+2-t_1-t_2)$, and where the indices are required 
to fulfill
\begin{equation*}
l_1+l_2+l_3=0\bmod 2\qquad \textrm{and}\qquad t_1+t_2+t_3=0\bmod 2 \, . 
\end{equation*}
The selection rule (\ref{mm0}) fixes the value of $s_3$ in terms of
$l_3$ and $t_3$.

The simple currents apart from the isomorphism class of the tensor unit are 
\begin{eqnarray} 
\Phi ^{0\ 0}_{k+1}&\sim &\Phi ^{k\ 1}_{\ 0}\qquad \textrm{for $k$ odd}\nn \\
\Phi ^{0\ 1}_{k+1}&\sim &\Phi ^{k\ 0}_{\ 0}\qquad \textrm{for $k$ even},\nn
\end{eqnarray}
hence $Pic(\cal{C})$ is cyclic of order two. The monodromy characters are 
products of monodromy characters for WZW theories based on $A_1$.

The only simple objects that can be used to apply Theorem \ref{ddc}
are $\Phi ^{1\ 1}_{\ 0}, \Phi ^{0\ 1}_{\ 1}$ and
$\Phi ^{0\ 1}_{\ k}$ for odd level $k$ and
$\Phi ^{1\ 0}_{k+1}$ for even level $k$.
The tensor product with the dual object is 
\begin{eqnarray}
\Phi ^{1\ 1}_{\ 0}\otimes \Phi ^{1\ 1}_{\ 0}&\cong&
\Phi ^{0\ 0}_{\ 0}\oplus \Phi ^{2\ 0}_{\ 0}\qquad \textrm{for level} \ k\ge 2\nn\\
\Phi ^{0\ 1}_{\ 1}\otimes\Phi ^{0\ 1}_{\ 1}&\cong &
\Phi ^{0\ 0}_{\ 0}\oplus\Phi ^{0\ 0}_{\ 2}\nn\\
\Phi ^{0\ 1}_{\ k}\otimes\Phi ^{0\ 1}_{\ k}&\cong &
\Phi ^{0\ 0}_{\ 0}\oplus\Phi ^{0\ 0}_{\ 2}\qquad \textrm{for odd level}\nn\\
\Phi ^{1\ 0}_{k+1}\otimes\Phi ^{1\ 0}_{k+1}&\cong&
\Phi ^{0\ 0}_{\ 0}\oplus\Phi ^{2\ 0}_{\ 0}\qquad \textrm{for even level}.\nn
\end{eqnarray}

$\Phi ^{0\ 0}_{\ 2}$ is a simple current only for $k=1$;
this case is indeed the Ising model with its well-known Kramers-Wannier
duality.  The primary field $\Phi ^{2\ 0}_{\ 0}$ is a simple current only for
$k=2$. In this case, we recover the known Kramers-Wannier duality of the 
tricritical Ising model.

\subsection{$N=1$ super-Virasoro minimal models}

The $N=1$ superconformal minimal models have the following coset description 
\cite{goko2} based on the diagonal embedding
\begin{equation}
(A_1)_{k+2} \hookrightarrow (A_1)_k\oplus (A_1)_2
\ . \nn
\end{equation}
To describe isomorphism classes of simple objects, consider triples
$\Phi ^{l\ s}_{\ t}$ with where $l\in \{0,\dots k\},\ s\in \{0,1,2\}$ and 
$t\in \{0, \dots k+2\}$, subject to the requirement
\begin{equation*}
\label{mm1}
l+s-t=0\bmod2.
\end{equation*}
For odd level $k$, representatives for simple objects are labelled
by equivalence classes of triples under the equivalence relation
\begin{equation*}
\Phi ^{l\ s}_{\ t}\,\, \sim \Phi ^{k-l \ 2-s}_{\ k+2-t}.
\end{equation*}
For even level, the same holds with the exception that there are
two simple objects corresponding to the triple 
$\Phi ^{\frac{k}{2}\ 1}_{\frac{k}{2}+1}$.
This phenomenon is called ``fixed point resolution'' in the physics literature
(see e.g.\ \cite{scya5} or \cite{scya6} for a review).

The tensor products with no fixed points involved are given by 
\begin{equation*}
\Phi ^{l_1\ s_1}_{\ t_1}\otimes \Phi ^{l_2\ s_2}_{\ t_2} \,\, \cong \,\,
\bigoplus_{l_3=|l_1-l_2|}^{l_{\max}}\ \ \bigoplus_{s_3=|s_1-s_2|}
^{s_{\max}}\ \ \bigoplus_{t_3=|t_1-t_2|}^{t_{\max}}
\Phi ^{l_3\ s_3}_{\ t_3}\, ,
\end{equation*}
where the indices are required to obey 
\begin{eqnarray}
l_1+l_2+l_3&=&0\bmod 2\nn\\
s_1+s_2+s_3&=&0\bmod 2\nn\\
t_1+t_2+t_3&=&0\bmod 2 \nn
\end{eqnarray}
with $l_{\max}=\min(l_1+l_2, 2k-l_1-l_2)$, 
$s_{\max}=\min(s_1+s_2, 4-s_1-s_2)$ and
$t_{\max}=\min(t_1+t_2, 2k+4-t_1-t_2)$. 

The simple currents are 
\begin{eqnarray} 
\Phi ^{0\ 0}_{\ 0}&\sim&\Phi ^{k\ 2}_{k+2}\nn\\
\Phi ^{0\ 2}_{\ 0}&\sim&\Phi ^{k\ 0}_{k+2}\nn\\
\Phi ^{0\ 0}_{k+2}&\sim&\Phi ^{k\ 2}_{\ 0}\qquad \textrm{for $k$ even}\nn \\
\Phi ^{0\ 2}_{k+2}&\sim&\Phi ^{k\ 0}_{\ 0}\qquad \textrm{for $k$ even},\nn
\end{eqnarray}
hence $Pic(\cal{C})\cong$ $\mathbb{Z}_2$ for odd level $k$ and
$Pic(\cal{C})\cong$ $\mathbb{Z}_2\times\mathbb{Z}_2$ for even level $k$.

We start with the discussion of dualities with cyclic stabilizer of order 
two. The duality classes have to be fixed points of quantum dimension $\sqrt2$ 
under the action of a nontrivial simple current of order two.
The quantum dimension 
of $\Phi ^{l\ m}_{\ s}$ is a product of the $A_1$ quantum dimensions.

We consider first fixed points of the simple current $\Phi ^{0\ 2}_{\ 0}$: 
in this case $s=1$ so that this $A_1$ summand already contributes
a multiplicative factor $\sqrt2$ to the quantum dimension; the other labels 
$l,t$ have to correspond to $A_1$-simple currents:
$l\in\{0,k\}$ and $t\in\{0,k+2\}$. This is excluded by the
selection rule $l+s+t=0\bmod 2$ for even level $k$. For odd level
$k$ we find a single duality class $\Phi ^{0\ 1}_{k+2}$ with
a cyclic stabilizer of order two generated by $\Phi ^{0\ 2}_{\ 0}$.

For even $k$ we have to consider fixed points under the action of
the simple current $\Phi ^{0\ 0}_{\ k+2}$
as well. In this case, the relevant $A_1$ constituent has level
greater or equal to three so that already this part makes a multiplicative
contribution to the quantum dimensions strictly bigger than $\sqrt2$. 
Thus, the cyclic group of order two generated by $\Phi ^{0\ 0}_{\ k+2}$ for 
$k$ even never occurs as a stabilizer.

We finally consider fixed points of the simple current $\Phi ^{k\ 0}_{\ 0}$
for even level $k$. By the
same arguments, quantum dimension $\sqrt2$ for a fixed point 
can only be achieved for $k=2$ and for $\Phi ^{1\ s}_{\ t}$ with $s,t$ 
describing simple currents. Such fixed points are, however, excluded by
the parity rule $l+m+s=0\bmod 2$.

For even level $k$, the full Picard group could appear as a
stabilizer as well. We should therefore find all fixed points 
of quantum dimension two under the action of all four simple currents.
Only the two simple objects arising in the ``fixed point resolution'' of
$\Phi ^{k/2\ 1}_{\ (k+2)/2}$ qualify. Indeed, a computation with
{\sc kac} \cite{kac} shows that quantum dimension 2 is achieved for level
$k=2$. Since the monotonicity lemma for WZW theories implies, by the
coset construction, the same monotonicity properties in $k$, this
is the only relevant case. A computation of the fusion rules,
e.g.\ again with {\sc kac}, however shows that the two simple
objects arising in the fixed resolution are not fixed points of
all four simple currents. Hence, for even $k$, there is no duality
object whose stabilizer is the full Picard group.

\subsection{$N=2$ super-Virasoro minimal models}

A coset realization for $N=2$ superconformal minimal models is based on
the embedding
\begin{equation}
u(1)_{2(k+2)}  \hookrightarrow (A_1)_k\oplus u(1)_4
\nn \, , 
\end{equation}
where $(u_1)_N$ stands for the modular tensor category with
$K_0((u(1)_N)= \mathbb{Z} \, / \, N \mathbb{Z}$ (for 
details, see e.g.\ Section 2.5.1 of \cite{fuRs4}).
To describe simple objects, consider triples $\Phi ^{l\ s}_{\ t}$
with $l\in \{0,\dots k\},\ s\in \mathbb{Z}/4\mathbb{Z}$ 
and $t\in \mathbb{Z}/2(k+2) \mathbb{Z}$, subject to the condition
\begin{equation*}
\label{mm2}
l+s-t=0\bmod2 \,\, .
\end{equation*}
Isomorphism classes of simple objects can be labelled by equivalence
classes of the equivalence relation
\begin{equation*}
\Phi ^{l\ s}_{\ t}\sim \Phi ^{k-l\ s\pm2}_{t\pm(k+2)}.
\end{equation*}
The tensor products read
\begin{equation*}
\Phi ^{l_1\ s_1}_{\ t_1}\otimes \Phi ^{l_2\ s_2}_{\ t_2} \,\, \cong \,\,
\bigoplus_{l_3=|l_1-l_2|}^{l_{\max}}\ \ 
\Phi ^{l_3\ (s_1+s_2)}_{\ (t_1+t_2)}
\end{equation*}
with $l_{\max}=\min(l_1+l_2, 2k-l_1-l_2)$, where the index $l$ must obey the 
selection rule
\begin{eqnarray}
l_1+l_2+l_3&=&0\bmod 2\nn \,\, . 
\end{eqnarray}

We can apply Theorem \ref{ddc} to the isomorphism class of simple objects
$\Phi ^{1\ 1}_{\ 0}$ for which the relevant tensor product
reads for $k\geq 2$:
\begin{equation*}
\label{fumi}
\Phi ^{1\ 1}_{\ 0} \otimes \Phi ^{1\ -1}_{0} \, \cong \,
\Phi ^{0\ 0}_{\ 0}\oplus \Phi ^{2\ 0}_{\ 0}\,.
\end{equation*}
Unless $k=2$, the second simple object is not a simple current.
For $k=2$ we find 16 simple currents and 8 duality classes 
$\Phi ^{1,\ \cdot}_{\ \cdot}$
which
all have a cyclic stabilizer generated by $\Phi ^{2\ 0}_{\ 0}$,
the worldsheet supercurrent.

The $N=2$ superconformal theory with $k=1$ has Virasoro central
charge $c=1$ and is equivalent to a free boson compactified on
a circle of appropriate radius. As a consequence,  all
isomorphism classes of simple objects are simple currents in this case and there
are no duality objects.

\vskip4em

\noindent {\bf Acknowledgements} \\[1pt]
We are grateful to J\"urgen Fuchs, Terry Gannon and Ingo Runkel for useful 
discussions. \\
E.T.\ is supported by a scholarship of the University 
of Hamburg (HmbNFG) and the GIF-Foundation. C.S.\ and E.T.\ received partial 
support from the Collaborative Research Centre 676 ``Particles, Strings and the
Early Universe - the Structure of Matter and Space-Time''.


 \def\Bi              {\bibitem}
 \newcommand\J[5]     {{\sl #5\/}, {#1} {#2} ({#3}) {#4} }
 \newcommand\K[6]     {{\sl #6\/}, {#1} {#2} ({#3}) {#4} \,{[#5]}}
 \newcommand\Prep[2]  {{\sl #2\/}, pre\-print {#1}}
 \newcommand\BOOK[4]  {{\sl #1\/} ({#2}, {#3} {#4})}
 \newcommand\inBO[7]  {{\sl #7\/}, in:\ {\sl #1}, {#2}\ ({#3}, {#4} {#5}), p.\ {#6}}
 \newcommand\inBOq[7] {{\sl #7\/}  in:\ {\sl #1} {#2}\ ({#3}, {#4} {#5}), p.\ {#6}}
 \newcommand\wb{\,\linebreak[0]} \def\wB {$\,$\wb}

\def\jf            {J.\ Fuchs}
\def\cfts          {conformal field theories}
\def\q             {quantum }
\def\Q             {Quantum }
\def\twodim        {two-di\-men\-sio\-nal}

 \def\adma  {Adv.\wb Math.}
 \def\apcs  {Applied\wB Cate\-go\-rical\wB Struc\-tures} 
 \def\anma  {Ann.\wb Math.}
 \def\anop  {Ann.\wb Phys.}
 \def\aspm  {Adv.\wb Stu\-dies\wB in\wB Pure\wB Math.}
 \def\aste  {Ast\'e\-ris\-que}
 \def\cocm  {Com\-mun.\wb Con\-temp.\wb Math.}
 \def\coma  {Con\-temp.\wb Math.}
 \def\comp  {Com\-mun.\wb Math.\wb Phys.}
 \def\cpma  {Com\-pos.\wb Math.}
 \newcommand\dgmd[2] {\inBO{{\rm Proceedings of the} XXth International Conference on
           Differential Geometric Methods in Theoretical Physics} {S.\ Catto and A.\ Rocha,
            eds.} \WS\Si{1991} {{#1}}{{#2}}}
 \def\fiic  {Fields\wB Institute\wB Commun.}
 \def\foph  {Fortschritte\wB d.\wb Phys.}
 \def\ijmp  {Int.\wb J.\wb Mod.\wb Phys.\ A}
 \def\ijmb  {Int.\wb J.\wb Mod.\wb Phys.\ B}
 \def\ihes  {Publ.\wb Math.\wB I.H.E.S.}
 \def\inma  {Invent.\wb math.}
 \def\jgap  {J.\wb Geom.\wB and\wB Phys.}
 \def\jhep  {J.\wb High\wB Energy\wB Phys.}
 \def\jlms  {J.\wB London\wB Math.\wb Soc.}
 \def\joal  {J.\wB Al\-ge\-bra}
 \def\jomp  {J.\wb Math.\wb Phys.}
 \def\jopa  {J.\wb Phys.\ A}
 \def\josp  {J.\wb Stat.\wb Phys.}
 \def\jpaa  {J.\wB Pure\wB Appl.\wb Alg.}
 \def\lemp  {Lett.\wb Math.\wb Phys.}
 \def\maan  {Math.\wb Annal.}
 \def\mams  {Memoirs\wB Amer.\wb Math.\wb Soc.}
 \def\mpla  {Mod.\wb Phys.\wb Lett.\ A}
 \newcommand\ncmp[2] {\inBO{IXth International Congress on
            Mathematical Physics} {B.\ Simon, A.\ Truman, and I.M.\ Davis,
            eds.} {Adam Hilger}{Bristol}{1989} {{#1}}{{#2}} } 
 \def\nuci  {Nuovo\wB Cim.}
 \def\nupb  {Nucl.\wb Phys.\ B} 
 \def\phlb  {Phys.\wb Lett.\ B}
 \newcommand\phgt[2] {\inBO{Physics, Geometry, and Topology}
            {H.C.\ Lee, ed.} \PL\NY{1990} {{#1}}{{#2}} }
 \def\phrb  {Phys.\wb Rev.\ B}
 \def\phrd  {Phys.\wb Rev.\ D}
 \def\phrl  {Phys.\wb Rev.\wb Lett.}
 \def\phrp  {Phys.\wb Rep.}
 \def\phrv  {Phys.\wb Rev.}
 \def\pnas  {Proc.\wb Natl.\wb Acad.\wb Sci.\wb USA}
 \def\ptrs  {Phil.\wb Trans.\wb Roy.\wb Soc.\wB Lon\-don}
 \newcommand\qfsm[2] {\inBO{\Q Fields and Strings: A Course for Mathematicians} 
            {P.\ Deligne et al., eds.} \AMS\PR{1999} {{#1}}{{#2}}}
 \def\rvmp  {Rev.\wb Math.\wb Phys.}
 \def\taac  {Theory\wB and\wB Appl.\wB Cat.}
 \def\tams  {Trans.\wb Amer.\wb Math.\wb Soc.}
 \newcommand\tgqf [2] {\inBO{Topology, Geometry and Quantum Field Theory
            {\rm [London Math.\ Soc.\ Lecture Note Series \# 308]}}
            {U.\ Tillmann, ed.} \CUP\Ca{2004} {{#1}}{{#2}}}
 \newcommand\tgqfq [2] {\inBOq{Topology, Geometry and Quantum Field Theory
            {\rm [London Math.\ Soc.\ Lecture Note Series \# 308]}}
            {U.\ Tillmann, ed.} \CUP\Ca{2004} {{#1}}{{#2}}}
 \def\trgr  {Trans\-form.\wB Groups}
 \def\AMS    {{American Mathematical Society}}
 \def\AP     {{Academic Press}}
 \def\BIR    {{Birk\-h\"au\-ser}}
 \def\CUP    {{Cambridge University Press}}
 \def\EMS    {{European Mathematical Society}}
 \def\IPC    {{International Press Company}}
 \def\NH     {{North Holland Publishing Company}}
 \def\KLU    {{Kluwer Academic Publishers}}
 \def\PUP    {{Princeton University Press}}
 \def\PL     {{Plenum Press}}
 \def\SV     {{Sprin\-ger Ver\-lag}}
 \def\WI     {{Wiley Interscience}}
 \def\WS     {{World Scientific}}
 \def\Ad     {{Amsterdam}}
 \def\Be     {{Berlin}}
 \def\Bo     {{Boston}}
 \def\Ca     {{Cambridge}}
 \def\Do     {{Dordrecht}}
 \def\pR     {{Princeton}}
 \def\PR     {{Providence}}
 \def\Si     {{Singapore}}
 \def\NY     {{New York}}

\end{document}